\documentclass[reprint,aps,superscriptaddress,amsmath,amssymb,longbibliography]{revtex4-1}

\usepackage{graphicx}
\usepackage{dcolumn}
\usepackage{bm}
\usepackage{color}
\usepackage{xspace}
\usepackage{ulem}
\usepackage{xfrac}
\usepackage[colorlinks=true,urlcolor=blue,citecolor=blue,linkcolor=blue,bookmarks=false,
pdfstartview={FitH}]{hyperref}

\def\cm-1{cm$^{-1}$\,}

\begin{document}

\title{Anomalous magneto-elastic coupling in Au-doped BaFe$_2$As$_2$}

\author{S.-F. Wu}\email{sfwu@iphy.ac.cn}
\affiliation{Department of Physics and Astronomy, Rutgers University,
Piscataway, NJ 08854, USA}
\affiliation{Beijing National Laboratory for Condensed Matter
Physics, and Institute of Physics, Chinese Academy of Sciences,
Beijing 100190, China}
\affiliation{School of Physical Sciences, University of Chinese
Academy of Sciences, Beijing 100190, China}          
\author{ W.-L. Zhang}
\affiliation{Department of Physics and Astronomy, Rutgers University,
Piscataway, NJ 08854, USA}
\author{L. Li}
\affiliation{Materials Science \& Technology Division, Oak Ridge
National Laboratory, Oak Ridge, TN 37831}    
\author{H.-B. Cao}
\affiliation{Neutron Scattering Division, Oak Ridge National
Laboratory, Oak Ridge, TN 37831}   
\author{ H.-H. Kung}
\affiliation{Department of Physics and Astronomy, Rutgers University,
Piscataway, NJ 08854, USA}             
\author{A.~S.~Sefat}    
\affiliation{Materials Science \& Technology Division, Oak Ridge
National Laboratory, Oak Ridge, TN 37831}
\author{H. Ding}
\affiliation{Beijing National Laboratory for Condensed Matter
Physics, and Institute of Physics, Chinese Academy of Sciences,
Beijing 100190, China}
\affiliation{School of Physical Sciences, University of Chinese
Academy of Sciences, Beijing 100190, China}
\affiliation{Collaborative Innovation Center of Quantum Matter,
Beijing, China}
\author{P. Richard}\email{pierre.richard.qc@gmail.com}
\affiliation{Beijing National Laboratory for Condensed Matter
Physics, and Institute of Physics, Chinese Academy of Sciences,
Beijing 100190, China}
\affiliation{School of Physical Sciences, University of Chinese
Academy of Sciences, Beijing 100190, China}
\affiliation{Collaborative Innovation Center of Quantum Matter,
Beijing, China}
\author{G. Blumberg}\email{girsh@physics.rutgers.edu}
\affiliation{Department of Physics and Astronomy, Rutgers University,
Piscataway, NJ 08854, USA}
\affiliation{National Institute of Chemical Physics and Biophysics,
12618 Tallinn, Estonia}
\date{\today}

\begin{abstract}                                             
                                      
We used polarization-resolved Raman scattering to study 
magneto-elastic coupling in Ba(Fe$_{1-x}$Au$_{x}$)$_2$As$_2$ crystals
as a function of light Au-doping, materials for which temperatures of
the structural transition ($T_S$) and of the magnetic ordering
transition ($T_N$) split. We study the appearance of the $A_g$(As)
phonon intensity in the $XY$ scattering geometry that is very weak
just below $T_S$, but for which the intensity is significantly
enhanced below $T_N$. In addition, the $A_g$(As) phonon shows an
asymmetric line shape below $T_N$ and an anomalous linewidth
broadening upon Au-doping in the magnetic phase. We demonstrate that
the anomalous behavior of the $A_g$(As) phonon mode in the $XY$
scattering geometry can be consistently described by a Fano model
involving the $A_g$(As) phonon mode interacting with the $B_{2g}$
symmetry-like magnetic continuum in which the magneto-elastic
coupling constant is proportional to the magnetic order parameter.

\end{abstract}
                                           
                       
\maketitle

\section{Introduction}

It is widely accepted that the magnetic and electronic properties of
the Fe-based superconductors are very sensitive to the As height with
respect to the Fe plane, and to the $\textrm{Fe-As-Fe}$ bond angle of
the Fe-As
tetrahedra~\cite{Kuroki2009PRB,Lee2014NJP,Baledent2015PRL,Vildosola2008PRB,Bascones2009PRB,YinPRL2008,Yndurain2011EPL,Cruz2010PRL,Zhang2014PRL,Lee2008JPSJ,Zhao2008,KurokiPRB2009,Garbarino2011EPL}.
The $c$-axis vibration of the As atom corresponds to a fully
symmetric phonon mode ($A_{1g}$) that modulates these two parameters.
First-principles calculations show that the phonon mode frequencies
agree well with experiments when the Fe magnetic ordering is
included~
\cite{Yildirim2009PhysicaC,BoeriPRB2010,Zbiri2009PRB,ReznikPRB2009,Hahn2009PRB,Mittal2013PRB,Hahn2013PRB},
and theoretical investigations suggest that the electron-phonon
coupling constant is enhanced in the magnetic
state~\cite{Zbiri2009PRB,Yndurain2009PRB,HuangPRB82,Bascones2013PRB,Coh2016PRB}.
The peak position of the fully symmetric As phonon density-of-states
in the calculation with the collinear magnetic structure decreases by
23\% compared with non-magnetic
calculations~\cite{Yildirim2009PhysicaC}. Experimental results
suggest that the fully symmetric As mode plays a major role in the
electron-phonon coupling in
BaFe$_2$As$_2$~\cite{Rettig2015PhysRevLett114}.

\begin{figure}[t]   
\begin{center}
\includegraphics[width=\columnwidth]{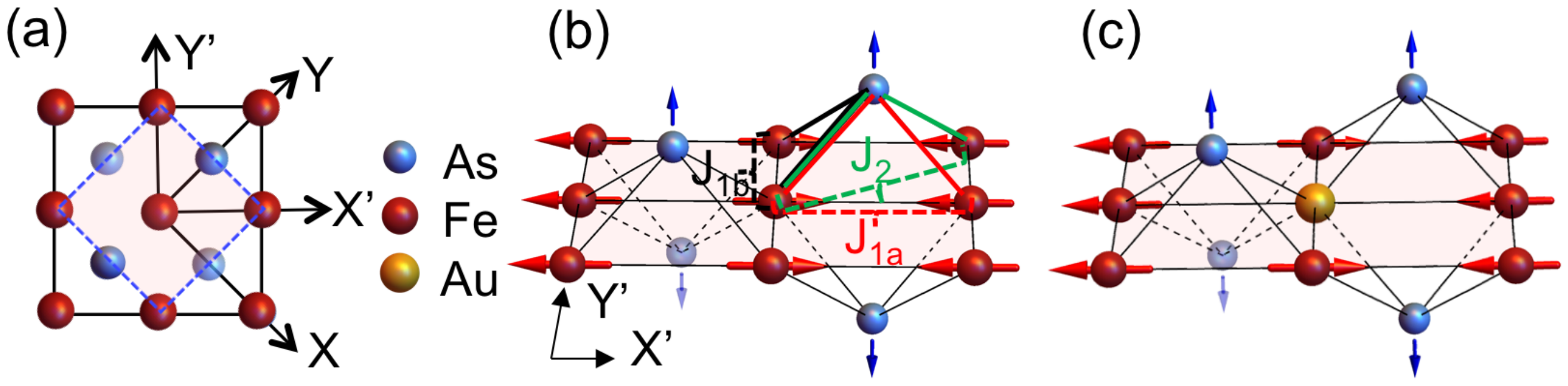}
\end{center}
\vspace{-6mm}
\caption{\label{Fig1_structure} 
(a) Definition of the $X$, $Y$, $X'$
and $Y'$ directions in tetragonal 2-Fe unit cell above $T_S$ (red
shaded area) and orthorhombic 4-Fe magnetic unit cell below $T_N$
(black solid lines). 
(b) Schematic diagram of the magnetic ordering
and of the fully symmetric As phonon mode. 
The red arrows mark Fe magnetic moments the collinear 
antiferromagnetic phase. 
The blue arrows indicate the As $c$-axes displacement. 
The red and black solid lines illustrate the super-exchange integrals of the
nearest Fe neighbors, $J_{1a}$ and $J_{1b}$. 
The blue solid lines
illustrate the super-exchange integral of the next-nearest Fe
neighbors, $J_{2}$.  
(c) Illustration of disorder effect by a non-magnetic Au in
the Fe plane of Ba(Fe$_{1-x}$Au$_{x}$)$_2$As$_2$.}    
\end{figure}  
Supported by previous Raman scattering
investigations~\cite{Choi2008PhysRevB78,ZhangWL2014arxiv,Chauvilere_PRB80,Kretzschmar2016NatPhy,Kaneko2017arxiv,Gnezdilov2013PRB87},
recently we reported a Raman study of a variety of parent compounds
of Fe-based superconductors in which we note a significant intensity
enhancement, below the N\'{e}el temperature $T_N$, of the emergent
fully symmetric As phonon mode measured in the unfavorable $XY$
scattering geometry~\cite{Ag_paper} (Fig.~\ref{Fig1_structure}). 
This enhancement is found only
for magnetically-ordered compounds. The results are interpreted in
terms of a magneto-elastic coupling. In agreement with previous Raman
studies on
Ba(Fe$_{1-x}$Co$_{x}$)$_2$As$_2$~\cite{Chauvilere_PRB80,Kretzschmar2016NatPhy},
we also observed an asymmetric line-shape for the fully symmetric As
mode below $T_N$ in the same $XY$ scattering geometry. Our results
are  described by a Fano model involving a coupling between the fully
symmetric As mode and the $B_{2g}$ electronic continuum, with the
coupling constant proportional to the magnetic order parameter
\cite{Ag_paper}. The purpose of this paper is to detail this model
and to apply it to Au-doped BaFe$_2$As$_2$, a system where the
structural phase transition temperature $T_S$ is a few Kelvin degrees
higher than the magnetic phase transition temperature
$T_N$~\cite{Li2015PRB,Au_susceptibility}. The experimental results
support the previous observations, confirm the validity of the model,
and gives about 1.5~meV as an estimate of the magneto-elastic
coupling constant in the parent compound. Interestingly, we find an
anomalous linewidth broadening upon Au-doping, essentially limited to
the magnetic phase.

\section{Experiment and Methods}\label{Experiment and Methods}

\begin{figure}[!b] 
\begin{center}
\includegraphics[width=\columnwidth]{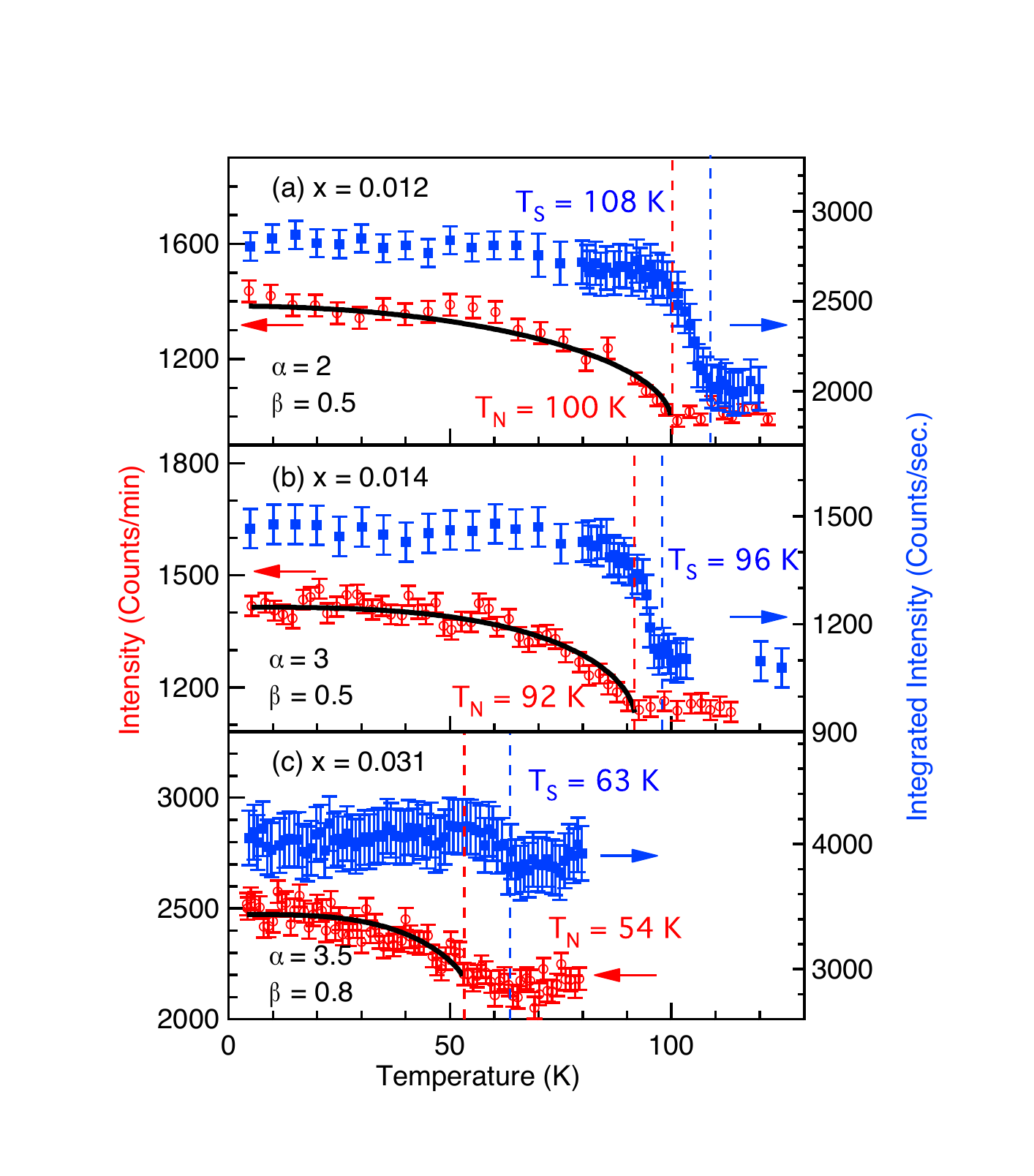}
\end{center}
\caption{\label{Fig2_fit_neutrons}  Neutron diffraction results for
Ba(Fe$_{1-x}$Au$_{x}$)$_2$As$_2$. (a) $x=0.012$, (b) $x=0.014$ and
(c) $x=0.031$. The blue solid squares represent the temperature
evolution, upon cooling, of the integrated intensity of lattice Bragg
peak $(2~2~0)$ in the tetragonal phase to $(4~0~0)$ in the
orthorhombic phase. 
The red open circles represent the temperature evolution, upon
cooling, of the intensity of magnetic Bragg peak
$($\sfrac{1}{2}$~$\sfrac{1}{2}$~5)$ in the
tetragonal phase to $(1~0~5)$ in the orthorhombic phase. The blue and
red dashed lines mark $T_S$ and $T_N$, respectively. The black solid
curves are the fits of the temperature evolution of the intensity of
magnetic Bragg peak $(1~0~5)$ with the formula
$I(T)=a[1-(T/T_N)^{\alpha}]^{2\beta}$.}
\end{figure} 
Single crystals of Ba(Fe$_{1-x}$Au$_{x}$)$_2$As$_2$~(x~=~0, 0.012,
0.014 and 0.031) were grown out of self-flux using a high-temperature
solution-growth technique described in
Refs.~\cite{Sefat2013bulk,Li2015PRB}, and the chemical compositions
were determined by inductive coupled plasma
analysis~\cite{Li2015PRB}. 
Neutron diffraction measurements on Au-doped samples were performed
using the four-circle diffractometer HB-3A at the High Flux Isotope
Reactor (HFIR) at the Oak Ridge National Laboratory to distinguish
the structural and magnetic transitions. A neutron wavelength of
1.542 \AA~was used from a bent perfect Si-220
monochromator~\cite{Chakoumakos2011JAC}.
The corresponding structural phase transition temperatures ($T_S$)
for Ba(Fe$_{1-x}$Au$_{x}$)$_2$As$_2$ are determined by the
temperature evolution of the integrated intensity of  lattice Bragg
peak $(2~2~0)$ in the tetragonal phase to $(4~0~0)$ in the
orthorhombic phase~[Fig.~\ref{Fig2_fit_neutrons}]. The corresponding
magnetic phase transition temperature ($T_N$) is determined for each
sample composition from the temperature evolution of the magnetic
Bragg peak intensities
$($\sfrac{1}{2}$~$\sfrac{1}{2}$~5)$ in the
tetragonal phase to $(1~0~5)$ in the orthorhombic
phase~[Fig.~\ref{Fig2_fit_neutrons}]. The $T_S$ and $T_N$ for the
parent compound BaFe$_2$As$_2$ are determined by resistivity and
magnetic susceptibility measurements~\cite{Li2015PRB}. All the $T_S$
and $T_N$ values for Ba(Fe$_{1-x}$Au$_{x}$)$_2$As$_2$ are summarized
in Table~\ref{TSTN}.

\begin{table}[!t]
\caption{\label{TSTN} Summary of the structural and magnetic phase
transition temperatures (Kelvin) for samples studied in this
manuscript. The last column is the ordered moment ($\mu_B$) per Fe at
4 K determined by neutron scattering measurements.}
\begin{ruledtabular}
\begin{tabular}{cccccc}
Sample&$T_{S}$&$T_{N}$&M\\
\hline  
BaFe$_2$As$_2$&135&135&0.87~\cite{Dai2015RevModPhys}\\
Ba(Fe$_{0.988}$Au$_{0.012}$)$_2$As$_2$&108&100&0.50$\pm$0.02\\
Ba(Fe$_{0.986}$Au$_{0.014}$)$_2$As$_2$&96&92&0.42$\pm$0.04\\
Ba(Fe$_{0.969}$Au$_{0.031}$)$_2$As$_2$&63&54&0.36$\pm$0.02\\
\hline
\end{tabular}
\end{ruledtabular}
\begin{raggedright}
\end{raggedright}
\end{table}    
The crystals used for Raman scattering were cleaved and positioned in
a continuous helium flow optical cryostat. The Raman measurements
were performed using the Kr$^+$ laser line at 647.1\,nm (1.92\,eV) in
a quasi-back scattering geometry along the crystallographic $c$-axis.
The excitation laser beam was focused into a $50\times100$ $\mu$m$^2$
spot on the $ab$-surface, with the incident power around 10~mW. The
scattered light was collected and analyzed by a triple-stage Raman
spectrometer, and recorded using a liquid nitrogen-cooled
charge-coupled detector. 
                             
The laser heating in the Raman experiments is determined by imaging
the appearance of stripes due to twin domain formation at the
structural phase transition temperature $T_S$
\cite{Kretzschmar2016NatPhy}. When stripes appear under laser
illumination, the spot temperature is just slightly below $T_S$, thus
$T_S=kP+T_{cryo}$, where $T_{cryo}$ is the temperature of cold helium
gas in the cryostat, $P$ is the laser power and $k$ is the heating
coefficient. By recording $T_{cryo}$ when the stripes appear at
different laser powers, we can deduce the heating coefficient using a
linear fit: $k =1 \pm0.1$ K/mW.

In this manuscript, we define the $X$ and $Y$ directions along the
two-Fe unit cell basis vectors (at 45$^{\circ}$ degrees from the
Fe-Fe directions) in the tetragonal phase, whereas $X'$ and $Y'$ are
along the Fe-Fe directions~[Fig.\ref{Fig1_structure}].

\section{Results and discussions}\label{Results}
    
\begin{figure*}[!t] 
\begin{center}
\includegraphics[width=\textwidth]{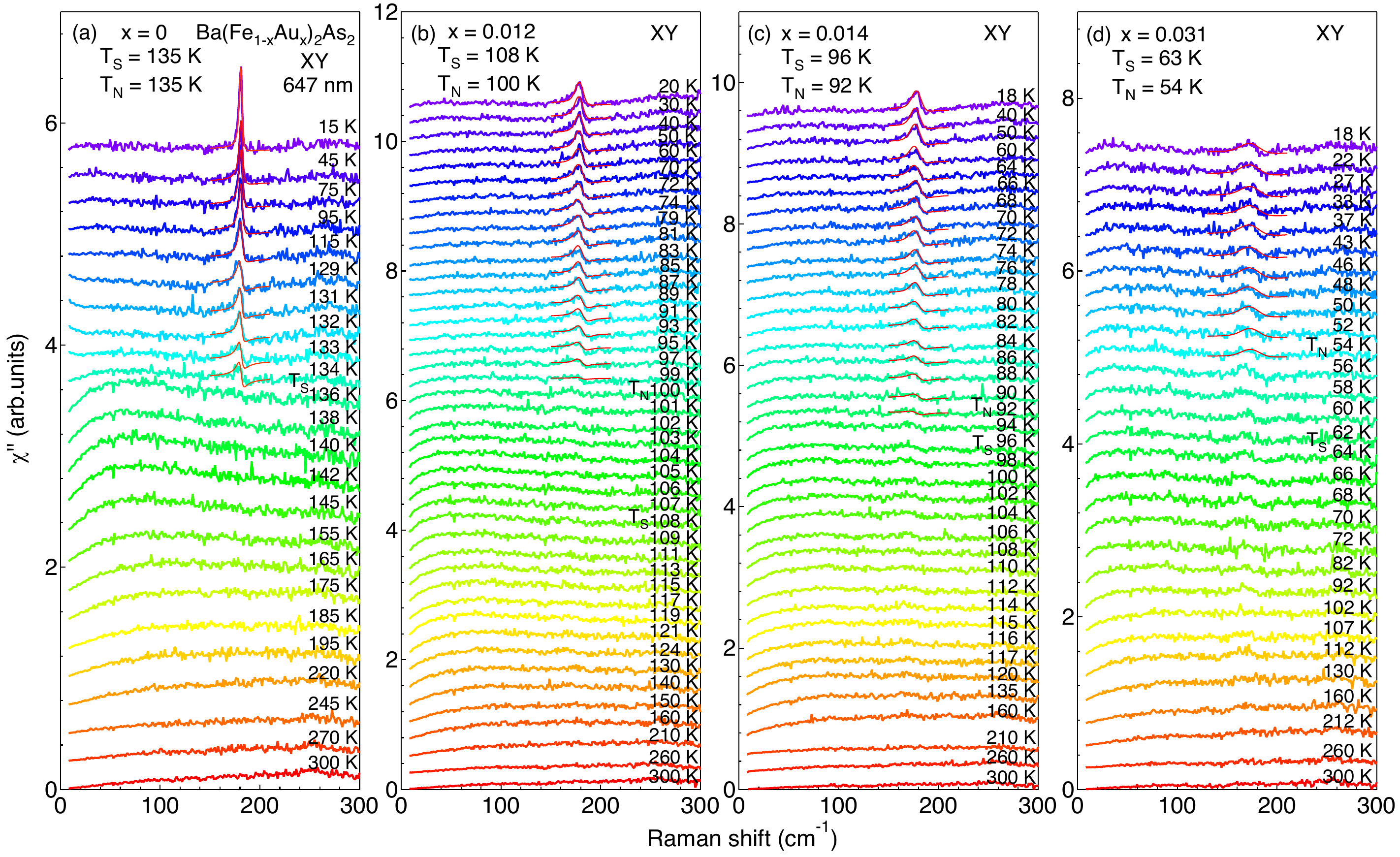}
\end{center}
\caption{\label{Fig3_XY} $T$-dependence of Raman spectra in the $XY$
scattering geometry in Ba(Fe$_{1-x}$Au$_x$)$_2$As$_2$. (a) $x=0$, (b)
$x=0.012$, (c) $x=0.014$, (d) $x=0.031$. The solid red line are fits
for the $A_g$(As) phonon mode using Eq.~\eqref{FanoShort}. The
spectral resolution is about 0.85 cm$^{-1}$.}
\end{figure*}   
\begin{table}[!b]
\caption{\label{symmetry} Summary of symmetry analysis in the
$D_{4h}$ and $D_{2h}$ point groups~\cite{CardonaBookII}.}
\begin{ruledtabular}
\begin{tabular}{ccccc}
Geometry&$D_{4h}$&$D_{2h}$\\
\hline
$XX$&$A_{1g}$+$B_{1g}$& $A_g$+ $B_{1g}$  \\
&&not a proper geometry\\
$XY$&$A_{2g}$+$B_{2g}$&$A_g+ B_{1g}$\\
&&not a proper geometry\\
$X'X'$&$A_{1g}$+$B_{2g}$&$A_g$\\
$X'Y'$&$A_{2g}$+$B_{1g}$&$B_{1g}$\\ 
$ZZ$&$A_{1g}$&$A_g$\\
\end{tabular}
\end{ruledtabular}
\begin{raggedright}
\end{raggedright}
\end{table}  
The body-centered crystal structure of BaFe$_2$As$_2$ in the high
temperature phase belongs to space group $I4/mmm$ (point group
$D_{4h}$). Below $T_S$, the space group symmetry lowers to $Fmma$
(point group $D_{2h}$). The non-degenerate Raman active phonon modes
at high-temperature ($D_{4h}$) are $A_{1g}$(As)+$B_{1g}$(Fe). The
Raman selection rules for the $D_{4h}$ point group indicate that the
$XX$, $XY$, $X'X'$ and $X'Y'$ scattering geometries probe $A_{1g} +
B_{1g}$, $A_{2g}+ B_{2g}$, $A_{1g} + B_{2g}$ and $A_{2g} + B_{1g}$,
respectively. However, in the orthorhombic phase with $D_{2h}$ point
group symmetry, the unit cell of BaFe$_{2}$As$_2$ rotates by 45
degrees; the $A_{1g}$ and $B_{2g}$ representations of the $D_{4h}$
point group merge into the $A_g$ representation of the $D_{2h}$ point
group, and $A_{2g}$ and $B_{1g}$ ($D_{4h}$) merge into $B_{1g}$
($D_{2h}$).

In the orthorhombic phase, the three proper scattering polarization
geometries are $X'X'$, $Y'Y'$ and $X'Y'$. Since the orthorhombicity is
small, the improper $XX$ and $XY$ polarizations still probe $A_g +
B_{1g}$ and $A_g$ symmetry excitations, respectively. The symmetry
correspondence for the point groups $D_{4h}$ and $D_{2h}$ are
summarized in Table~\ref{symmetry}~\cite{Zhang2016PRB}.

In Figs.~\ref{Fig3_XY}(a)-\ref{Fig3_XY}(d), we show detailed
temperature evolution of Raman spectra for
Ba(Fe$_{1-x}$Au$_x$)$_2$As$_2$ ($x=0$, 0.012, 0.014 and 0.031) in the
$XY$ scattering geometry. The $T_S$ and $T_N$ values of each sample,
carefully characterized by neutron
scattering~[Fig.~\ref{Fig2_fit_neutrons}], are indicated in each
panel. For the parent compound BaFe$_2$As$_2$
[Fig.~\ref{Fig3_XY}(a)], the $A_g$(As) phonon mode appears instantly
below 134 K, which is close to $T_S$ and $T_N$. The phonon mode
rapidly sharpens upon cooling and becomes more symmetric. For
Ba(Fe$_{0.988}$Au$_{0.012}$)$_2$As$_2$, $T_S$ (108~K) and $T_N$
(100~K) are split. The Raman data, displayed in
Fig.~\ref{Fig3_XY}(b), show that the $A_g$(As) mode is hardly
detectable between $T_S$ and $T_N$. Similarly as in Co-doped
BaFe$_2$As$_2$, the mode gains strength only below $T_N$
\cite{Kretzschmar_Nature12}. Although the asymmetry of the phonon
line shape decreases upon cooling, we note that it remains more
asymmetric than in pristine BaFe$_2$As$_2$ down to the lowest
temperature. We report similar observation for
Ba(Fe$_{0.986}$Au$_{0.014}$)$_2$As$_2$ [Fig.~\ref{Fig3_XY}(c)] and
Ba(Fe$_{0.969}$Au$_{0.031}$)$_2$As$_2$ [Fig.~\ref{Fig3_XY}(d)]
\footnote{We note that for Ba(Fe$_{0.969}$Au$_{0.031}$)$_2$As$_2$
$A_{1g}$(As) phonon intensity appears above $T_S$ in forbidden $XY$
polarization (see Fig. 3(d)). The leakage is due to the local
symmetry breakdown for crystals with high Au impurity concentration
for which the selection rules are relaxed.}, although the peak
becomes broader due to the disorder introduced by Au-doping. 
 
\begin{figure}[!t] 
\begin{center}
\includegraphics[width=\columnwidth]{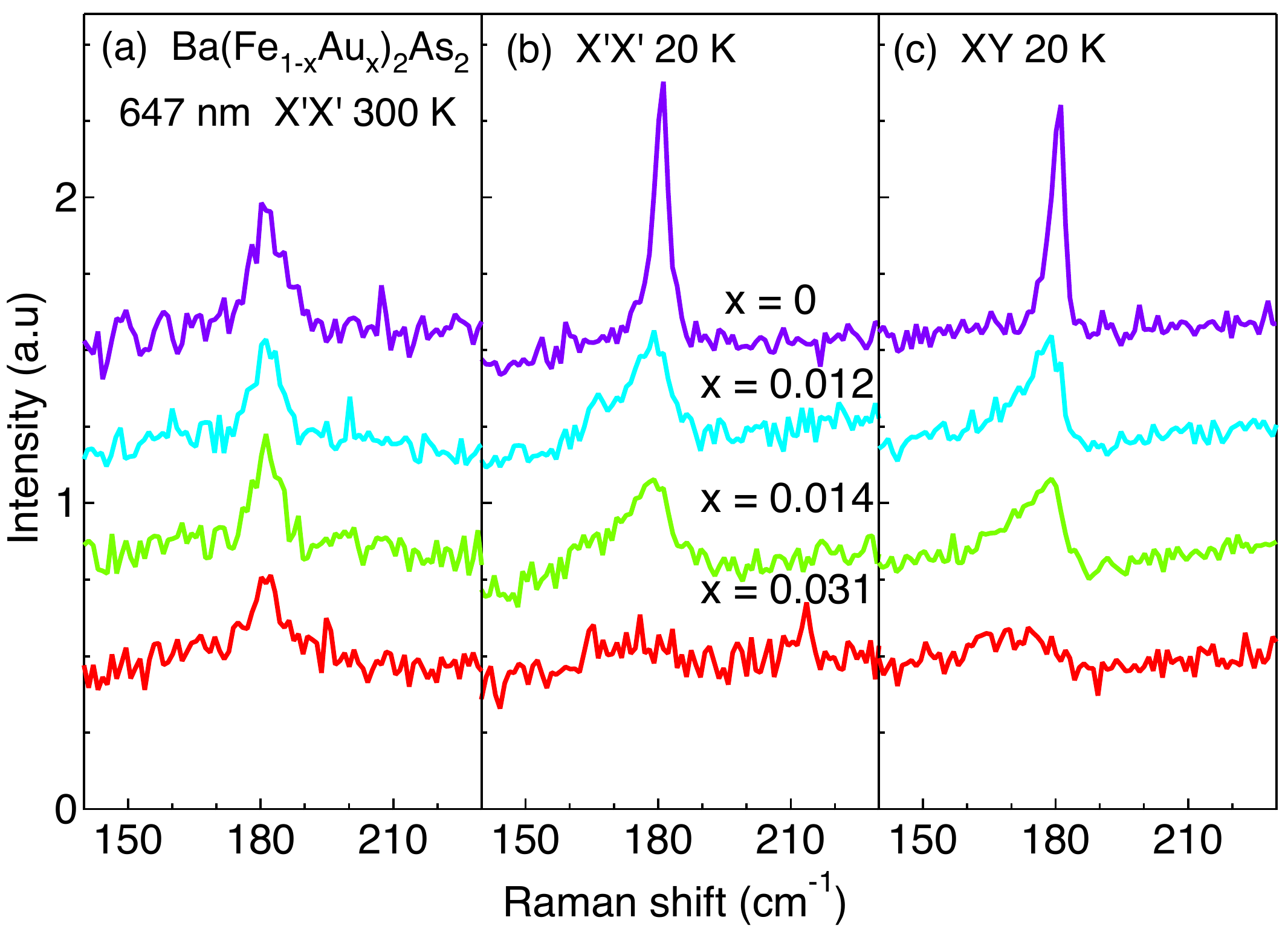}
\end{center}
\caption{\label{Fig4_XpXp_10K_300K} (a) Doping dependence of the
$A_{1g}$(As) phonon peak in Ba(Fe$_{1-x}$Au$_x$)$_2$As$_2$ in the
$X'X'$ scattering geometry at 300 K. (b) Same as (a) but at 20 K. (c)
Same as (b) but in the $XY$ scattering geometry.} 
\end{figure}
Broadening of the spectra due to the disorder introduced from dopants
is expected. However, Fig.~\ref{Fig4_XpXp_10K_300K} indicates that
the disorder effect in Ba(Fe$_{1-x}$Au$_x$)$_2$As$_2$ is far from
trivial. Indeed, the data show a broader line-width at 20~K than at
300~K. This would be expected only if disorder itself has a strong
impact on the magneto-elastic coupling. In
Figs.~\ref{Fig4_XpXp_10K_300K}(a) and \ref{Fig4_XpXp_10K_300K}(b), we
compare the doping evolution of the $A_{1g}/A_g$(As) phonon in the
$X'X'$ scattering geometry at 300 K and 20 K. At 300 K, in the
non-magnetic phase, the $A_{1g}$(As) phonon is symmetric and shows
nearly doping-independent mode frequency, linewidth and intensity, as
shown in Fig.~\ref{Fig4_XpXp_10K_300K}(a).
Fig.~\ref{Fig4_XpXp_10K_300K}(b) illustrates contrasting behavior for
the $A_g$(As) phonon at 20~K. We observe significant broadening,
weakening, softening and a pronounced asymmetric line-shape upon
Au-doping. Similar anomalous $A_g$(As) phonon behavior in the
magnetic phase is also detected in the $XY$ scattering geometry at
20~K [Fig.~\ref{Fig4_XpXp_10K_300K}(c)].

\begin{figure}[t] 
\begin{center}
\includegraphics[width=\columnwidth]{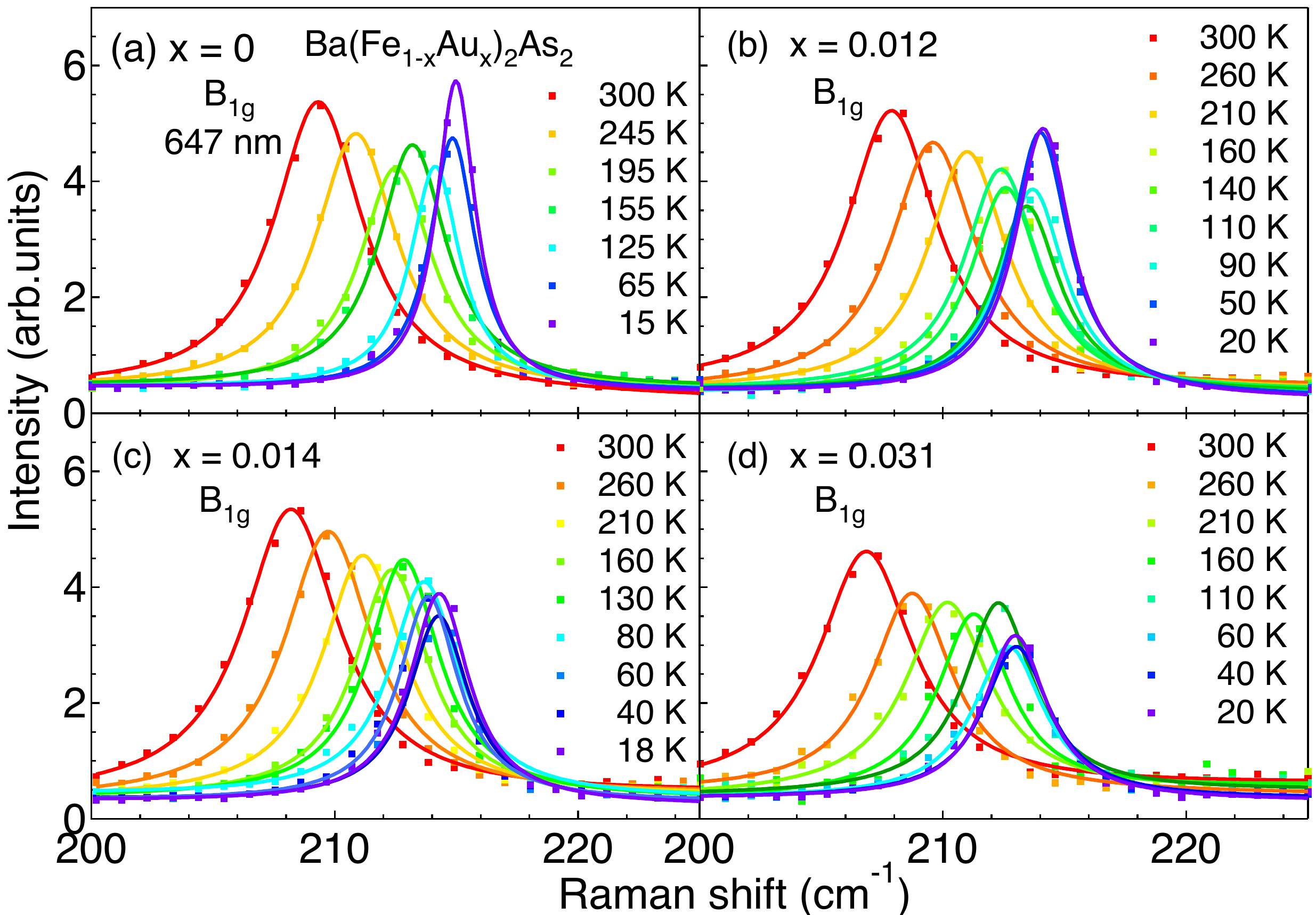}
\end{center}
\caption{\label{Fig5_B1g_4doping} $T$-dependence of Raman spectra of
the $B_{1g}$(Fe) phonon in the $X'Y'$ geometry for
Ba(Fe$_{1-x}$Au$_x$)$_2$As$_2$. (a) $x=0$, (b) $x=0.012$, (c)
$x=0.014$, (d) $x=0.031$. The solid line are the Lorentzian fits of
the $B_{1g}$(Fe) phonon peak at different temperatures.}
\end{figure}

Before moving forward with a more quantitative analysis of the
$A_g$(As) phonon, it is instructive to take a closer look at the
$B_{1g}$ phonon, associated with the vibration of the Fe atom along
the $c$ axis. While below the orthorhombic transition, the $A_{1g}$
phonon is symmetry-allowed to couple to the $B_{2g}$-like electronic
continuum, such coupling is not possible for the $B_{1g}$(Fe) phonon,
and accordingly, the line shape of this mode in the $X'Y'$
configuration remains symmetric for all temperatures. However,
\textit{a priori} the $B_{1g}$(Fe) phonon could couple to the
$B_{1g}$-like electronic continuum. To investigate this possibility,
we display in Fig. \ref{Fig5_B1g_4doping} the temperature evolution
of the $B_{1g}$(Fe) phonon at different Au-doping levels. 
                                 
\begin{figure}[t] 
\begin{center}
\includegraphics[width=\columnwidth]{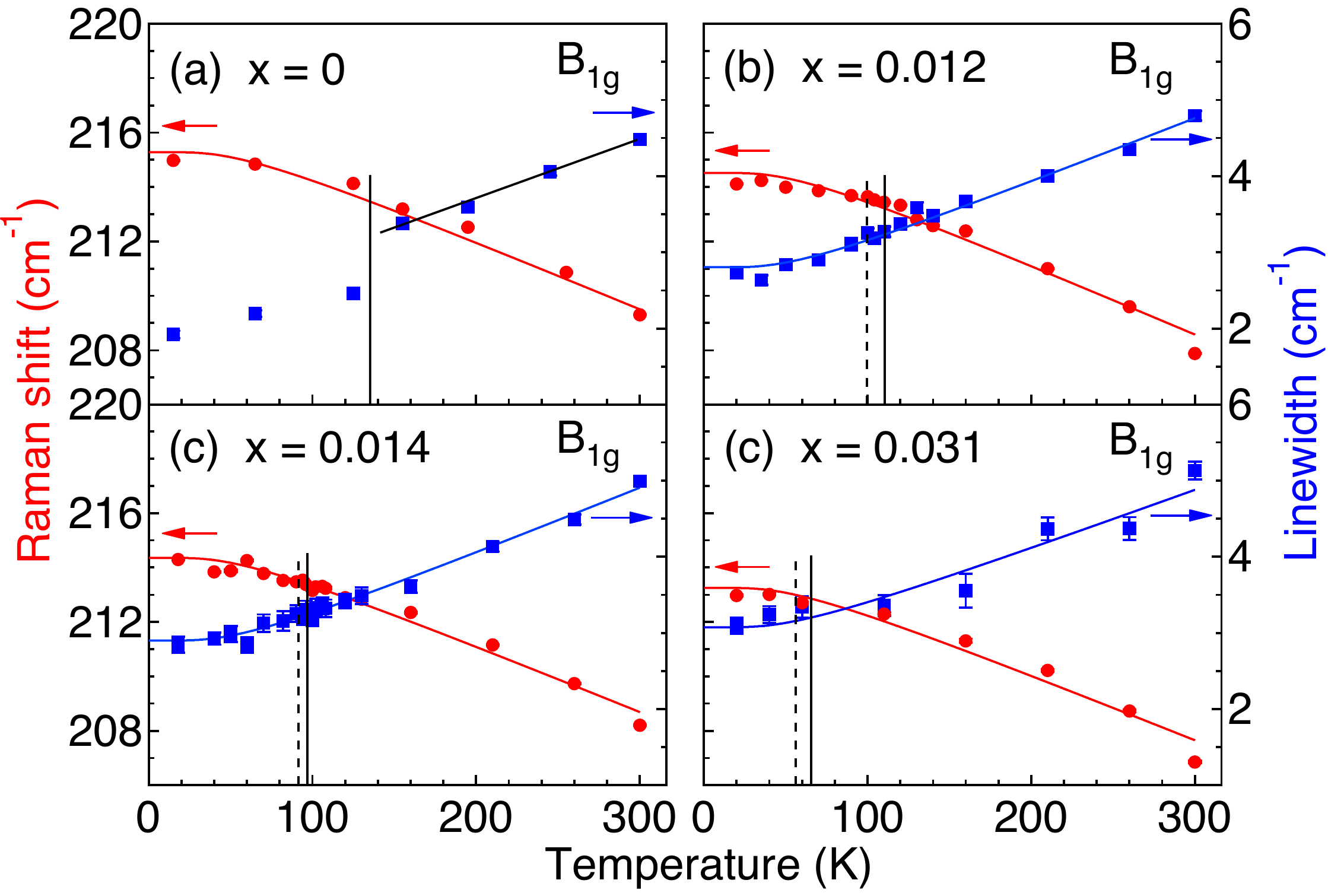}
\end{center}
\caption{\label{Fig6_B1g_width} $T$-dependence of the peak frequency
(red dots) and line-width (blue squares) for the $B_{1g}$(Fe) phonon
in Ba(Fe$_{1-x}$Au$_x$)$_2$As$_2$. (a) $x=0$, (b) $x=0.012$, (c)
$x=0.014$ and (d) $x=0.031$. The solid red lines are fits for the
$T$-dependence of the peak frequency using Eq.~\eqref{eq_omega},
whereas the solid blue line are fits for the $T$-dependence of the
linewidth using Eq.~\eqref{eq_gamma}. The vertical dashed lines
represent $T_N$. The vertical solid black lines correspond to $T_S$,
when different from $T_N$.}
\end{figure}

Upon cooling, the $B_{1g}$(Fe) phonon mode hardens and sharpens
without detected anomalies around $T_S/T_N$ (see
Figs.~\ref{Fig6_B1g_width}(a)-\ref{Fig6_B1g_width}(d)), except for
the linewidth of the parent compound BaFe$_2$As$_2$, which displays a
discontinuity around $T_S/T_N$~\cite{GallaisPRB2011}. The temperature
dependence of the mode frequency and linewidth for the four Au-doping
concentrations can be fitted by the anharmonic phonons decay
model~\cite{Klemens_PhysRev148,Menendez_PRB29}:
\begin{equation}
\label{eq_omega}
\omega_{ph}(T)=\omega_{0}-C\left( 1+\frac{2}{e^{\frac{\hbar\omega_0
}{ 2k_{B}T}} -1} \right )
\end{equation}
\begin{equation}
\label{eq_gamma}
\Gamma_{ph}(T)=\Gamma_{0}+\Gamma\left(
1+\frac{2}{e^{\frac{\hbar\omega_0 }{ 2k_{B}T}} -1} \right ).
\end{equation}
  \begin{table}[!b]
\caption{\label{B1g_Parameters} 
Summary of fitting parameter for
$B_{1g}$(Fe) phonon mode in Ba(Fe$_{1-x}$Au$_{x}$)$_2$As$_2$. }
\begin{ruledtabular}
\begin{tabular}{ccccc}
Sample&$\omega_{0}$&$C$&$\Gamma_{0}$&$\Gamma$~(\cm-1)\\
\hline
BaFe$_2$As$_2$&217.25&1.97&-&-\\
Ba(Fe$_{0.988}$Au$_{0.012}$)$_2$As$_2$&216.54&2.02&2.14&0.67\\
Ba(Fe$_{0.986}$Au$_{0.014}$)$_2$As$_2$&216.28&1.92&2.22&0.68\\
Ba(Fe$_{0.969}$Au$_{0.031}$)$_2$As$_2$&215.14&1.89&2.46&0.61\\
\end{tabular}
\end{ruledtabular}
\begin{raggedright}
\end{raggedright}
\end{table}
The fitting parameters are summarized in Table~\ref{B1g_Parameters}. 
The $B_{1g}$(Fe) phonon shows slight
softening and broadening upon Au doping. 
There is no indication of a
coupling between the $B_{1g}$(Fe) phonon and the $B_{1g}$ electronic
continuum. We also note that the linewidth of the $B_{1g}$(Fe) phonon
broadens only by 1~cm$^{-1}$ from $x=0$ to $x=0.031$ at 20~K,
demonstrating that Au-doping barely affects the $B_{1g}$(Fe) phonon,
in contrast to the $A_g$(As) phonon.

As discussed in Ref.~\cite{Ag_paper}, the enhancement of the
$A_g$(As) phonon intensity upon entering the magnetic phase is
related to the magnetic moment. Although there is increasing evidence
for a description of the magnetism of the Fe-based superconductors
beyond simple weak coupling or strong coupling theories
\cite{Bascones_CR_Physique36}, it worths taking a look at these
theories in order to get hints on a microscopic description of the
enhancement of the $A_g$(As) phonon intensity.

The enhancement of the $A_g$(As) phonon intensity can be rationalized
in several ways. A previous theoretical paper~\cite{Bascones2013PRB}
proposes two mechanisms to explain the enhancement of the $A_g$(As)
phonon intensity due to the electron-phonon coupling upon entering
the magnetic phase. The first mechanism is directly related to the
geometrical parameters such as the Fe-As-Fe angle, whereas the second
one has to do with the modification of the Fe-As Slater-Koster energy
integrals $pd\sigma$ and $pd\pi$ due to As displacements. Both
mechanisms result in finite intensity in the $B_{2g}$ channel when
the magnetic moment is finite, which is consistent with our
experimental results.

The $A_g$(As) phonon intensity enhancement can also be understood
from a $J_1-J_2$ model derived from the strong coupling approach
\cite{Si_PRL101,C_Fang_PRB77,CK_Xu_PRB78,CK_Xu_PRB78}. Here the
effective nearest and next-nearest neighbors super-exchange
parameters $J_1$ and $J_2$ are determined by the details of the
Fe-As-Fe configuration. Below $T_N$, the effective super-exchange
constant $J_1$ becomes anisotropic along the $X'$ and $Y'$
directions, either due to the anisotropy of the ordered magnetic
moment~\cite{Nevidomskyy2011arxiv}, or due to the difference in the
electron hopping probability along and perpendicular to the stripe
directions~\cite{Anderson1959PR}. As the $A_{g}$(As) phonon $c$-axes
vibration modulates the effective super-exchange parameters $J_{1a}$
and $ J_{1b}$ \textit{via} the As bridge, $|J_{1a}-J_{1b}|$ is also
modulated because the super-exchange Fe-As-Fe path along the $X'$ and
$Y'$ directions are different. The relative anisotropy of the
in-plane electronic polarizability along the $X'$ and $Y'$
directions, as induced by the $A_{1g}$(As) phonon, is proportional to
$|J_{1a}-J_{1b}|$. The $|J_{1a}-J_{1b}|$ term becomes nonzero only
below $T_N$ when the collinear AFM order is established, explaining
why the $A_g$(As) phonon in the $XY$ geometry appears below $T_N$
with enhanced intensity.

Whether it originates from the local mechanism
described just above, or it is caused by the bi-quadratic interaction
revealed to be large due to deviations from a local picture
\cite{Wysocki_NPHYS7}, the effective $|J_{1a} - J_{1b}|$ term,
proportional to the in-plane electronic polarizability, was shown to
be proportional to the square of the magnetic ordered moment
$M$~\cite{Nevidomskyy2011arxiv}. Furthermore, the $A_g$(As) phonon
intensity ratio in $XY$ and $XX$ scattering geometries $I_{XY}/
I_{XX}$ is also demonstrated to scale with the square of the magnetic
ordered moment $M$~\cite{Ag_paper}. These results suggest a
magneto-elastic coupling with a strength proportional to the magnetic
order parameter~\cite{Han2009PRL,
Rettig2015PhysRevLett114,Mandal2014PRB}, a fact that we are going to
exploit.

In order to quantify the electron-phonon interaction evidenced by the
Raman data in the $XY$ scattering geometry below $T_N$, we introduce
a Fano model in which the $A_g$(As) phonon couples to the
$B_{2g}$-like electronic continuum~\cite{Ag_paper}. The resulting
line shape, which describes the interference between a discrete phonon
mode and an interacting
continuum~\cite{Fano1961PhysRev,CardonaBookI,blumberg1994JSC}, has
the following form: 
\begin{equation}
\label{FanoShort}
I(\omega)=T_{e}^2 \frac{\pi \rho
(\omega_0-\omega-v\frac{T_{ph}}{T_{e}})^2}{(\omega_0 - \omega)^2 +
(v^2 \pi \rho )^2}   
\end{equation}
where $\omega_0$ is the bare phonon frequency, $v$ is the
magneto-elastic coupling constant, $T_{ph}$ and $T_{e}$ are the Raman
coupling amplitudes to the phonon and to the electronic continuum,
respectively, and $\rho$ is the electronic density-of-states in the
vicinity of the phonon frequency.

\begin{figure}[!t] 
\begin{center}
\includegraphics[width=\columnwidth]{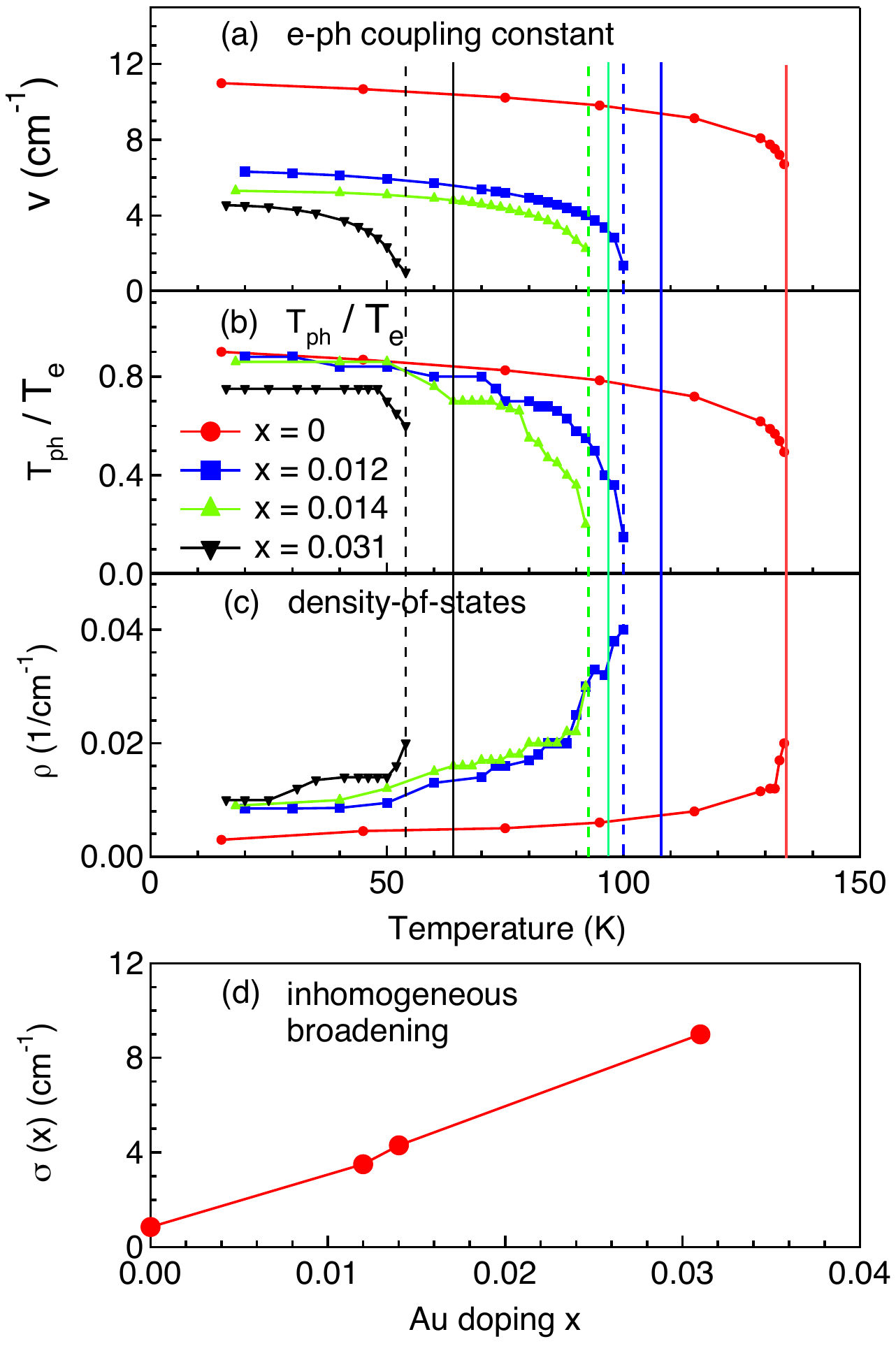}
\end{center}
\caption{\label{Fig7_vbr} $T$-dependence of the fitting parameters.
(a) Electron-phonon coupling constant $v(T)$. (b) Ratio
$T_{ph}/T_{e}(T)$ of the phonon transition matrix element and the
electronic continuum transition matrix element. (c) Electronic
density-of-states $\rho(T)$. (d) Inhomogeneous broadening factor
$\sigma(x)$. $\sigma(0)$ is due to instrumental resolution. 
The solid lines and the dashed lines mark the
corresponding $T_S$ and $T_N$ for the four dopings, respectively. }
\end{figure}

For BaFe$_2$As$_2$, we derive the temperature dependence of the
magnetic order parameter $M(T)=b(1-T/135)^{0.103}$ obtained from the
fitting of the temperature evolution of the $(1~0~3)$ magnetic Bragg
peak intensity \cite{Wilson2009PRB} (the magnetic Bragg peak
intensity is proportional to the square of the magnetic order
parameter). Since $\omega_0$ barely changes upon cooling, we use the
constant $\omega_0=181.4$ cm$^{-1}$ derived from the peak frequency
in the $XX$ scattering geometry at 15~K. We also fix the coupling of
light to the electronic continuum $T_{e}^2=1.4$, and we keep
$T_{ph}/T_{e}$ and $\rho$ as two $T$-dependent fitting parameters.
The spectrum at 15~K is well reproduced with $T_{ph}/T_{e}=0.9$,
$\rho=0.006$ states/cm$^{-1}$, and $v=11$ cm$^{-1}$. The temperature
dependence of $v(T)$, which is proportional to the magnetic order
parameter, is shown in Fig.~\ref{Fig7_vbr}(a). The temperature
evolutions of $T_{ph}/T_{e}(T)$ and $\rho(T)$ are illustrated in
Figs.~\ref{Fig7_vbr}(b) and \ref{Fig7_vbr}(c), respectively.

\begin{table*}[t]
\caption{\label{Parameters} Summary of fixed parameters.}
\begin{ruledtabular}
\begin{tabular}{ccccccc}
Au-doping 
$x$&$M(T)$&$v(T)$~(\cm-1)&$\omega_0$~(\cm-1)&$T_e^2$&$\sigma(x)$~(\cm-1)\\
\hline
0&$d[1-(T/135)]^{0.103}$&
$11[1-(T/135)]^{0.103}$&181.4&1.4&0.85\\
0.012&$e[1-(T/100)^{2}]^{0.25}$&
$6.3[1-(T/100)^{2}]^{0.25}$&179.3&1.4&3.5\\
0.014&$f[1-(T/92)^{3}]^{0.25}$&
$5.3[1-(T/92)^{3}]^{0.25}$&178&1.4&4.3\\
0.031&$g[1-(T/53)^{3.5}]^{0.4}$&
$4.5[1-(T/53)^{3.5}]^{0.4}$&175&1.4&9\\
\end{tabular}
\end{ruledtabular}
\begin{raggedright}
\end{raggedright}
\end{table*}

For Au-doped BaFe$_2$As$_2$ below $T_N$, we convolute the Fano
intensity $I(\omega)$ with a Gaussian inhomogeneous broadening factor
$\sigma(x)$. The $\sigma(x)$ broadening is determined by fitting the
spectra at the lowest temperature for each doping. The $\sigma(x)$
values are summarized in Fig.~\ref{Fig7_vbr}(d).

The temperature dependence of the $(1~0~5)$ Bragg peak intensity
below $T_N$ is fitted with the formula
$I(T)=a[1-(T/T_N)^{\alpha}]^{2\beta}$, as shown in
Fig.~\ref{Fig2_fit_neutrons}. Thus, the normalized magnetic order
parameter $M(T)$ is obtained using
$M(T)=c[1-(T/T_N)^{\alpha}]^{\beta}$
[Table~\ref{Parameters}].
As shown in Figs. \ref{Fig3_XY}(b)-\ref{Fig3_XY}(d), the Raman data
for the Au-doped samples are well described by the model. The
corresponding temperature dependence of $T_{ph}/T_{e}(T)$ and
$\rho(T)$ for all Au dopings are given in Figs.~\ref{Fig7_vbr}(b) and
\ref{Fig7_vbr}(c), respectively. The sets of fixed parameters are
given in Table~\ref{Parameters}. We note that we have normalized the
parameters $v$ and $\frac{T_{ph}}{T_{e}}$ to their values at 20 K. We
also kept the electronic continuum transition matrix elements $T_e$
the same for all Au-doping concentrations.

As shown in Fig.~\ref{Fig7_vbr}(b), the Raman coupling amplitude to
the phonon ($T_{ph}$) increases upon cooling as an order parameter.
This observation is reasonable because $T_{ph}$ itself should be
proportional to the lattice orthorhombicity $\delta$.

The electronic density-of-states represented by the parameter
$\rho(T)$ decreases upon cooling, which is consistent with the
opening of a spin-density-wave gap
\cite{Hu2008PRL,RichardPRL2010,Zhang2016PRB,Chauvilere_PRB80}. This
observation provides a natural explanation for the reduction of the
asymmetry in the $A_g$(As) phonon line shape upon cooling below $T_N$
\cite{Chauvilere_PRB80}.

The parameter $\sigma(x)$ describes the inhomogeneous broadening
effect due to the Au substitution of Fe. As shown in
Fig. \ref{Fig1_structure}(c), the non-magnetic Au disorder introduced
at the Fe position significantly perturbs the local magnetic order
and weakens the Fe-Fe spin-spin correlation length. Indeed, only
3.1\% Au-doping is sufficient to lower $T_N$ from 135~K to 53~K. Each
non-magnetic local Au dopant has four As neighbors and effectively
changes the coupling strength between the $A_{1g}$(As) mode and the
local Fe magnetic order parameter. The Au-doping has a major effect
on the $A_g$(As) phonon, unlike the $B_{1g}$(Fe) phonon. The
broadening of the $A_g$(As) phonon at $x=0.031$ is 9~cm$^{-1}$, while
it is only 1 cm$^{-1}$ at 20~K for the $B_{1g}$(Fe) phonon. The
significant broadening of the $A_g$(As) phonon mode upon Au-doping in
the magnetic state indicates a strong coupling of the $A_g$(As)
phonon to magnetism~\cite{Bascones2013PRB}.

Finally, the enhanced electron-phonon coupling in the collinear AFM
phase may have implications to superconductivity in the Fe-based
superconductors. Earlier calculations of the electron-phonon coupling
constant $\lambda$ in the nonmagnetic phase led to
$\lambda=0.2$~\cite{Boeri2008PRL101}. According to recent
calculations, the electron-phonon matrix element is enhanced four
times in the collinear AFM phase due to the presence of
$d_{xz}/d_{yz}$ Fermi surfaces around the $M$
point~\cite{Coh2016PRB,Coh2015NJP}. The enhancement is notably
important at 22 meV in the Eliashberg $\alpha^2F$ spectral function,
which coincides with the energy of the $A_{1g}$ mode. Therefore, the
intra-band paring temperature is possibly  enhanced due to the
electron-phonon coupling.

\section{Conclusion}

In conclusion, we used polarized Raman scattering to study Au-doped
BaFe$_2$As$_2$ samples for which the $T_S$ and $T_N$ transition
temperatures split. Our results confirm that the intensity of the
$A_g$(As) mode in the $XY$ scattering geometry is
enhanced only below $T_N$ and also reveal an asymmetric phonon
line shape of the $A_g$(As) mode that becomes more symmetric upon
cooling. To describe the line shape of the phonon peaks, we adopted a
Fano model accounting for the magneto-elastic coupling. The
enhancement of intensity below $T_N$ is consistent with a
magneto-elastic coupling constant proportional to the magnetic order
parameter, allowing to transfer the apparent phononic Raman intensity
from the electronic continuum~\cite{Klein1984}. The magneto-elastic
coupling constant, estimated to about 1.5~meV in the
parent compound, is non-negligible. The temperature dependence of the
line shape asymmetry is explained by the interference of the coherent
light scattering amplitudes due to the As phonon mode and the
interacting $XY$-symmetry continuum, the intensity of which
monotonically diminishes upon cooling as the spin-density-wave gap
opens when the magnetic order develops.                
                                                            	
We also note that when the magnetic Fe atom is substituted by
non-magnetic Au at a few per cent level, the $B_{1g}$(Fe) phonon
associated with the Fe $c$-axis vibration shows only little
disorder-induced broadening. In contrast, the anomalous $A_g$
symmetry As mode appearing in the $XY$ scattering geometry below
$T_N$ shows significant enhancement of the inhomogeneous broadening
with Au-doping, underlying the magneto-elastic coupling mechanism.
The inferred inhomogeneous broadening of the anomalous $A_g$(As) mode
reaches 1 meV at 3.1\% gold substitution.

The pronounced Fano line-shape and significant broadening of the
$A_g$(As) phonon mode upon Au-doping in the magnetic state
demonstrates the strong coupling of the $A_g$(As) phonon to magnetism
and to the electronic continuum in the Fe-based superconductors, with
possible consequences on the intra-band paring temperature.
      
\begin{acknowledgments}
We thank E. Bascones and K. Haule for discussions.
The research at Rutgers was supported by the US Department of Energy,
Basic Energy Sciences, and Division of Materials Sciences and
Engineering under Grant No. DE-SC0005463. The work at ORNL was
supported by the US Department of Energy, Basic Energy Sciences,
Materials Sciences and Engineering Division. Work at IOP was
supported  by grants  from NSFC (11674371 and 11274362) and  MOST
(2015CB921301, 2016YFA0401000 and 2016YFA0300300) of China.
\end{acknowledgments}

\end{document}